\documentstyle[12pt,aas2pp4,psfig]{article}
\newcommand{\asca}{{\it ASCA}}
\newcommand{\chandra}{{\it Chandra}}
\newcommand{\axj}{G0.570$-$0.018}

\begin{document}

\title{Unusual diffuse X-ray source in the Galactic center region} 

\author{Atsushi~Senda, 
Hiroshi~Murakami, and  Katsuji~Koyama} 

\affil{Department of Physics, Graduate School of Science, Kyoto University,
Sakyo-ku, Kyoto 606-8502, Japan; senda@cr.scphys.kyoto-u.ac.jp,
hiro@cr.scphys.kyoto-u.ac.jp, koyama@cr.scphys.kyoto-u.ac.jp}
\begin{abstract}

We report the {\asca} and {\chandra} discovery of a diffuse X-ray source
in the Galactic center region. The X-ray spectrum is fitted with a non-equilibrium ionization (NEI) 
plasma model of about 6-keV temperature.  
The model requires higher than solar metal abundances,
a young plasma age of $\simeq$ 100 years and a large $N_{\rm H}$ value of about 10$^{23}$ cm$^{-2}$.  
The $N_{\rm H}$  value constrains the source position to be in 
the Galactic center region at about 8.5~kpc distance.  
The high resolution X-ray image with the {\chandra} ACIS 
shows a ring of 10$''$ radius which corresponds to 0.4 pc at the 
Galactic center, and a tail-like structure.
Although the morphology is peculiar, the other X-ray features
are likely to be a very young supernova remnant, possibly
in a free expansion phase.   
 
\end{abstract}

\keywords{Galaxy: center --- ISM: individual (G0.570$-$0.018/CXO J174702.6$-$282733 ) --- Supernova remnants --- X-rays: ISM }

\section{Introduction}
The Galactic center (GC) region within  $\simeq$100~pc is 
very rich in various X-ray sources, such as X-ray binaries, 
young stellar clusters and supernova remnants (SNRs).  
A big mystery is a diffuse hot ($\simeq$ 10 keV) plasma of 
$1\arcdeg \times 2\arcdeg$~elliptical shape with the total energy of 
about $10^{54}$ ergs (Koyama et al. 1989, 1996;  Yamauchi et al. 1990).  
The dynamical age of the hot plasma is estimated to be about $10^5$~years.
One possible origin of the diffuse plasma is 
multiple supernova explosions (10$^{3}$~supernovae) 
in the past 10$^5$~years (Yamauchi et al. 1990), 
which predicts that many young SNRs should be discovered with a deep exposure 
above the 2 keV X-ray band.
With the {\asca}~Galactic center survey project, we found an extended 
X-ray source on the Galactic plane near the giant molecular cloud Sgr~B2,    
which exhibits unusually strong iron lines (Sakano et al. 1998).  
The {\chandra}~Cycle~1 observation confirmed the {\asca}~results and
revealed a ring and tail structures in the iron K$_{\alpha}$~line band, 
which we named G0.570$-$0.018/CXO J174702.6$-$282733 (here and after, {\axj}). 
This paper reports the {\asca} and {\chandra} results of {\axj} and 
discuss the nature of this peculiar source.

\section{Observations and Data Reductions}

{\asca} (Tanaka et al. 1994) observed the Sgr B2 region near
the Galactic center on October 1 in 1993 (PV phase)  
and September 22$-$24 in 1994 (AO2).
{\asca} was equipped with two Gas Imaging Spectrometers (GIS2 and GIS3) 
and two Solid-state Imaging Spectrometers (SIS0 and SIS1) 
on the focal planes of four X-ray telescopes (XRT) 
(Selemitsos et al. 1995; Ohashi et al. 1996; Burke et al. 1991).
In both the observations, {\axj} was placed on the center of the  
SIS-field, hence was intercepted by the center gap of 
each CCD chip.  We therefore used the GIS (GIS2 and GIS3) data
only.  For GISs, nominal bit-assignment of PH-mode wes used in 
high and medium bit rate. After the standard data screening, 
we obtained the effective exposure time of PV and AO2  
of 17.5~ksec and 80.0~ksec, respectively.

{\axj} was in the ACIS-I field of view when {\chandra}
observed  the Sgr~B2 region on 29$-$30 March 2000.
The satellite and instruments are described by Weisskopf et al. (1996) and 
Garmire et al. (2001), respectively.  Data reduction is made with  
the same method of the Sgr~B2 analysis (Murakami, Koyama, \& Maeda 2001).
{\axj} is located at the top of chip I3, and thus is heavily affected 
by the Charge Transfer Inefficiency (CTI).
To minimize the effect of the CTI degradation, we use the software
developed by Townsley et al. (2000).
We correct the energy gain using the instrumental emission 
line from  Ni (K$_{\alpha}$=7.5keV) and Au atoms (L$_{\alpha}$=9.7keV)
with an accuracy of  $<0.5$\% (90\% confidence). 
The response matrix is made using the nearly contemporaneous observation 
of reference lines from an on-board calibration source (OBSID$=$62097), 
which are analyzed with the same CTI correction. The effective area of 
the telescope mirrors and the detection efficiency of ACIS are calculated 
with the {\it mkarf} program in the {\chandra} {\it Interactive Analysis of 
Observations Software} (CIAO, Version 2.1). After screening the data, 
we obtain the effective exposure time of 100~ksec.

\section{Results and Analysis}

\subsection{X-ray Morphology}

In the {\asca} GIS image pointed at the Sgr~B2 cloud, we find a local 
excess in the west of Sgr~B2.  This excess is clearly seen if we limit 
the X-ray energy band of 6.0$-$7.0~keV, which include the iron K-shell 
transition line.  
Figure~1 shows the 6.0$-$7.0~keV  band image.
The brightest source located at the northeast is the giant molecular cloud  
Sgr~B2 (Murakami et al. 2000).  The other excess at the center of the image 
is a newly discovered X-ray source.
Due to the limited statistics and spatial resolution, error of the 
peak position of the diffuse structure is rather large of 
$0\fdg 55 \lesssim l \lesssim 0\fdg 57$, $ - 0\fdg 02 \lesssim b \lesssim 0\fdg 00$.

In order to investigate  the accurate position and fine spatial structure of 
this diffuse source,
we inspect the  {\chandra} image in the 4.0$-$7.0 keV band.
As is shown in Figure~2, we see a complex X-ray structure with a ring of about 
10$''$ radius  and an east-to-west tail from the ring.
The center coordinate of the ring structure is 
R.A.$=$~17$^{\rm h}$47$^{\rm m}$02$\fs$6, 
Dec$=$~$-$28$\arcdeg$27$\arcmin$33$\arcsec$ (epoch 2000), 
which corresponds to ($l$, $b$) = (0$\fdg$570, $-$0$\fdg$018), 
the same position of the {\asca} diffuse source  within the error. 
We hence designate this source
as G0.570$-$0.018/CXO J174702.6$-$282733 (hereafter {\axj}).  
To investigate the statistical significance of the shell-like structure,
 we make a radial profile with the center at (17$^{\rm h}$47$^{\rm m}$02$^{\rm s}$.6, $-$28$\arcdeg$27$\arcmin$33$\arcsec$)$_{\rm{J}2000}$.
Figure~3 shows the radial profile in the 4.0$-$7.0~keV band.  
The excess around 10$''$ is statistically significant, hence 
the ring-structure is real.
On the other hand, a faint point-like structure at the shell center 
seen in Figure~2 is not significant because it contains only 3 photons.
Unlike Sgr~B2, no radio nor any other wave band counterpart
is found for {\axj} (see Oka et al. 1998a, 1998b;
Tsuboi, Handa, \& Ukita 1999; Lis{,} \& Carlstrom 1994; 
Mehlinger et al. 1992, 1993; Seiradakis et al. 1989). 

\subsection{X-ray Spectra}

The {\asca} X-ray spectrum is made using the data in a circle of 3-arcmin
radius and subtracting the background data in a 3$'$ $< r <$ 4$'$~ ring 
as is shown in Figure 1.
The spectrum exhibits prominent line at 6.0$-$7.0~keV and large absorption 
at low energy band. We then fit with a phenomenological model of a thermal 
bremsstrahlung plus a Gaussian line.
The line energy is determined to be $6.60_{-0.09}^{+0.12}$~keV 
with the equivalent width of $3.7_{-1.2}^{+3.0}$~keV 
(here and after, errors are $90\%$ confidence unless otherwise noted), 
indicating a $K_\alpha$ line from iron atoms with 
lower ionization states than helium-like (6.7~keV).
The temperature is constrained to be higher than 2~keV. Therefore
the equilibrium  ionization state of iron should be helium or hydrogen 
like with the line energy of 6.7$-$6.9~keV.  
We thus conclude that the plasma is still in an ionizing phase or in 
a non-equilibrium ionization (NEI).

The {\chandra} X-ray spectrum is made from the same area as with {\asca}, 
while the background spectrum is taken  from a source-free
region with the same CTI effect as the source region.
The background-subtracted flux is 20400 counts and the spectrum shows a  
clear iron line as is already found with {\asca}.
In the diffuse source region, 
we also find many faint point-like sources. We hence execute the 
CIAO '{\it wavdetect}' software of a wavelet method 
(Freeman et al. 2000) and resolve 18 point sources.
The integrated flux (2.0$-$10.0 keV) of all the point sources is
410 counts, which is only $\simeq$ 2\% of the diffuse X-rays. 
Therefore a contamination of the point-sources can be ignored practically.
The {\chandra} spectrum is well fitted with the same phenomenological model 
as used for the {\asca} spectrum, a bremsstrahlung plus a Gaussian line.   
The plasma temperature is constrained to be $>3.4$ keV, while the line energy is $6.50_{-0.03}^{+0.03}$ keV with the equivalent width of $4.1_{-1.0}^{+1.4}$ keV. 
Thus the best-fit parameters are consistent with, and are more accurate 
than the {\asca} results.

Since both the {\asca} and {\chandra} spectra are fitted with the same  
model and the line center energy is consistent with an NEI plasma, 
we simultaneously fit the two spectra with a more realistic
NEI plasma model (Borkowski et al. 2001). 
In addition to a conventional thin thermal plasma model,
an NEI model includes another parameter, 
which is called the ionization parameter $\tau$=$nt$,
where $n$~and $t$~are plasma density and elapsed time after 
the plasma is heated-up.  
As the parameter $\tau$~becomes larger than 
10$^{12}$ $\sim$ 10$^{13}$~cm$^{-3}$ s, 
an NEI plasma approaches to a collisional ionization 
equilibrium (CIE) plasma.
In addition to the strong iron line, 
the {\chandra} spectrum includes emission lines at the energies of 
K-shell transition of argon, calcium atoms. 
We therefore vary the abundances of all the elements collectively 
fixing the relative ratio to be solar (Anders, \& Grevesse 1989).
Free parameters are the electron temperature ($kT$), the abundance ($Z$), 
the ionization parameter ($nt$), 
and interstellar absorption ($N_{\rm H}$). For the other free parameter,
we allow one normalization factor to be fitted both the {\asca}
and {\chandra} fluxes simultaneously. 
The NEI fit is acceptable with the best-fit spectra and parameters shown in
Figure~4 and Table~1.  The data excess of {\chandra} and deficit of {\asca} 
above and below the iron line are attributable to the gain uncertainty of 
these instruments.
However this small systematic difference is not a serious problem in 
the present analysis.   
The ionization parameter of $\tau$ =1.7$\times$10$^{10}$~cm$^{-3}$~s 
is very small, while the abundance is significantly 
larger than the solar value.   

\section{Discussion}
{\asca} discovered a diffuse  X-ray source ({\axj}) with a strong iron line 
near the Sgr B2 cloud,
then {\chandra} found a ring-like structure of 10$''$-radius.   
The hydrogen column density is determined to be 
$N_{\rm H}=(13.9_{-3.2}^{+3.3})\times 10^{22}$~H cm$^{-2}$. 
Since the $N_{\rm H}$~value is nearly equal to those of the Galactic center 
sources of the same Galactic lattitude (Sakano 2000), 
{\axj} would be located near at the Galactic center region. 
We hence assume the source distance to be 8.5~kpc.  
The X-ray luminosity is then estimated to be $\simeq 10^{34}$~ergs and the 
size of the ring radius is 0.4~pc. 
The spectra are well fitted with an NEI model with plasma temperature 
($kT$) of 6.1 keV, ionization parameter ($\tau$) of 
1.7$\times$10$^{10}$~cm$^{-3}$~s,
and metal abundances ($Z$) of 4.5 solar.
Therefore, together with the ring-like morphology, {\axj}~is likely
a young SNR, either in an adiabatic or a free expansion phase.  
We first apply a Sedov self-similar model assuming in 
an adiabatic phase.  
This model gives the emission measure ($E.M.= n^{2}V$), 
radius ($R$) and electron temperature ($kT$) as follows 
(Ostriker et al. 1988);\\

$n^{2}V = (4n_{a})^{2} \times 4\pi R^{2} \times (\frac{R}{12})$ \\
$(\rm{cm}^{-3})$,

$R = 5.0 \times (\frac{E}{10^{51}~\rm {ergs}})^{1/5}
(\frac{n_{a}}{1 ~\rm{cm}^{-3}})^{-1/5}(\frac{t_{s}}{10^{3}~\rm{years}})^{2/5}~$ \\
$(\rm {pc})$, and
 
$kT = 4.5 \times (\frac {E}{10^{51}~\rm {ergs}})^{2/5}
(\frac{n_{a}}{1~\rm {cm}^{-3}})^{-2/5}(\frac{t_{s}}{10^{3} ~\rm {years}})^{-6/5}$ \\
$(\rm {keV})$,\\

where $t_s$, ~$n_{a}$ and $E$ are the age of the plasma 
(in unit of year), 
ambient density (in units of cm$^{-3}$) and explosion energy of the 
supernova (in units of ergs), respectively.
Using the observed values of $R$ = 0.4 pc, $kT$ = 6.1 keV  and 
$E.M.$ = 8.1$\times$10$^{56}$ cm$^{-3}$,
we obtain $t_s$, $n_a$ and $E$ to be  70 years, 
5.1 cm$^{-3}$ and  3.5 $\times$ 10$^{48}$ ergs, 
respectively.   The swept-up  mass is then calculated to be 
$n_{a}\times 4/3\pi R^{3} \simeq 0.03 {\rm M_{\odot}}$, 
which is extremely smaller than that of the SN ejecta of a few 
${\rm M_{\odot}}$. 
The estimated explosion energy  of {\axj}~is also extremely small 
compared with a usual SNR. 
These  indicate that only  a tiny fraction of the explosion energy 
($\sim 10^{51}$~ergs) and ejected mass  (a few ${\rm M_{\odot}}$)  have 
been converted to the thermal plasma.  These facts strongly indicate that 
{\axj}~is not in an adiabatic phase but still in a free expansion phase. 

The  X-ray ring structure of {\axj}~is similar to that of SN 1987A 
(Burrows et al. 2000), where a strong stellar wind 
from a massive progenitor might produce a gas ring 
of sub-pc radius, and was heated by the collision of supernova ejecta.  
From the radial profile in Figure 3, the thickness of the 
ring (FWHM) is about $6''$, or 0.2~pc, hence the plasma volume is 
1.2$\times$10$^{55}$ cm$^3$. Since the X-ray flux from the ring is $67\%$ 
of that of the whole area of {\axj}, the emission measure ($E.M.$) from the 
ring is  5.5$\times$10$^{56}$ cm$^{-3}$. Then using the best-fit ionization 
parameter ($nt$) of 1.7$\times$10$^{10}$~cm$^{-3}$~s, 
the plasma density ($n$) and ionization age ($t$) of the X-ray ring 
are estimated to be 6.7 cm$^{-3}$  and 80 year, respectively. 
In Table 2, we compare the physical parameters of the ring of {\axj} 
with those of SN 1987A.
The ring size and  age are about 3$-$6 times larger 
than those of SN 1987A. Therefore {\axj}~would be a "future SN 1987A" 
after a free expansion with the expected speed of 
$v = R/t \simeq$ 4900 km s$^{-1}$. 
   
One problem of this scenario for {\axj} is an additional X-ray structure, 
the east-to-west X-ray tail.  We suspect that the structure would be 
made by some instability of dense stellar wind from progenitor, and 
was heated-up by the collision of the SN ejecta.  In this case, the front 
of free expanding ejecta would be about two times larger than 
the ring radius, hence the age of SNR is doubled. This age is still 
in a very young phase, thus the above discussion based on a free expansion 
phase is essentially not changed.   

In the young SNR scenario, we expect non-thermal radio emissions associated
 with the thermal X-rays. However we see no significant radio continuum flux 
from {\axj}.
In general, the X-ray flux is proportional to the square of electron 
density ($n^2$),
while the radio flux is proportional to the square of magnetic field, $B^2$.
Assuming that magnetic field is proportional to the ambient gas density  
(the magnetic field is frozen to the ambient gas), the flux ratio between 
X-ray to radio is roughly constant among shell-like SNRs. 
Referring the radio and X-ray data of typical young shell-like SNRs,  
Cas A, Tycho and Kepler, we estimated the radio flux of {\axj} to be  
0.1$-$1 Jy (except Cas A) at 1 GHz, using the X-ray flux of $\sim$10$^{-12}$ 
ergs s$^{-1}$. 
SN 1987A has almost the same X-ray flux as that of {\axj}, and the radio flux density
is $\sim$40 mJy at 4.7 GHz (Burrows et al. 2000).  
These radio flux density would be below the current detection limit of 
the radio SNRs located near the GC, where the radio background is extremely 
high ($\sim$0.1$-$0.5 Jy). 
Thus we encourage more deep and fine spatial resolution search for radio 
emissions  near at {\axj}.

We express our sincere thanks to T. Tsuru, and M. Kohno 
for their contributions in very early phase of this study and 
stimulating discussions on this paper. 
The authors also thank Y. Maeda and M. Sakano 
for their useful discussion.
H. Murakami is financially supported by JSPS grant No 199904648.
 
\newpage

\bibliography{apj-jour}

\newpage
\onecolumn
\begin{figure}
\centering \leavevmode
\psfig{file=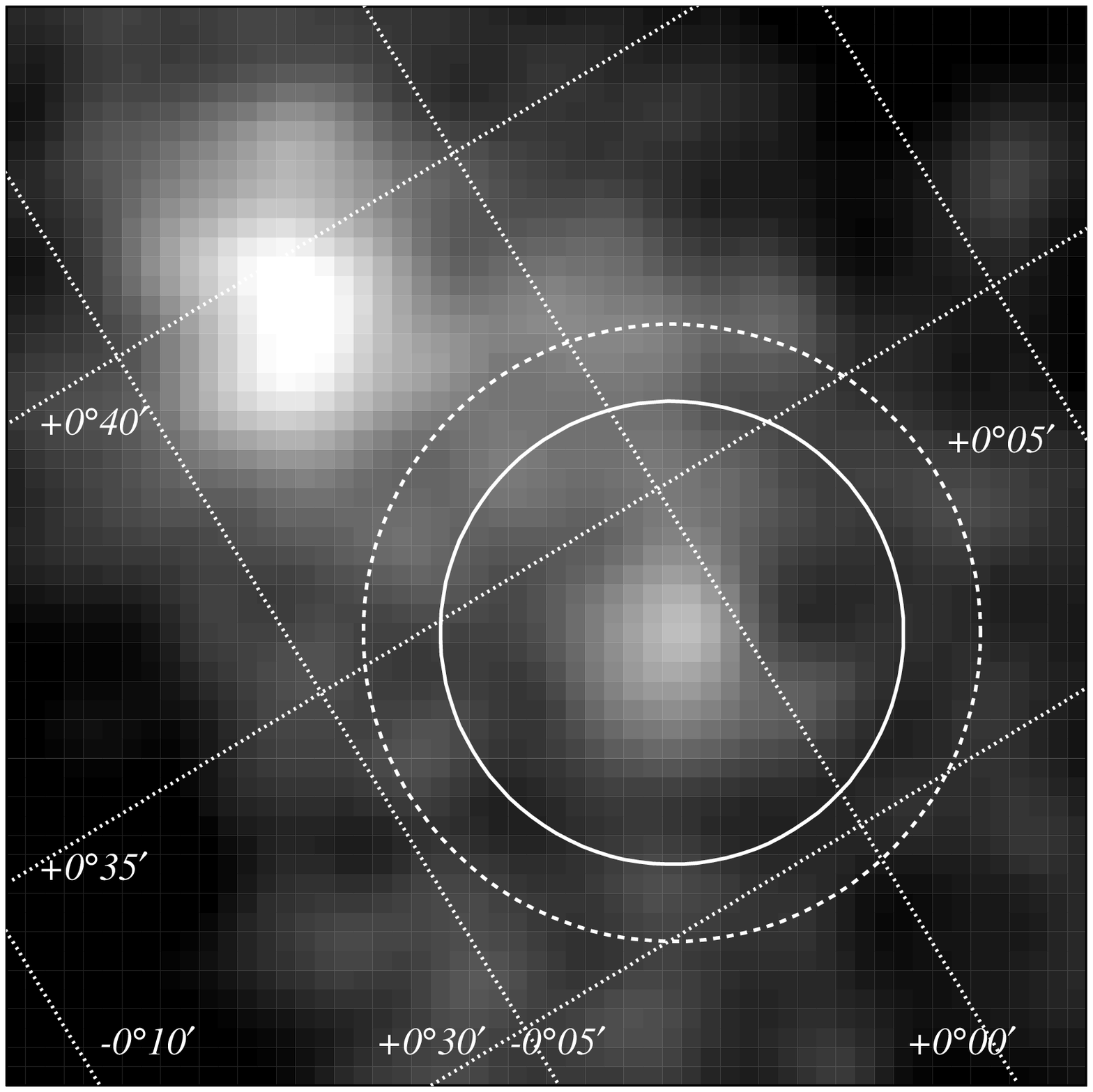,width=0.8\textwidth}
\figcaption[f1.eps]{{\asca}~GIS image in the 6.0$-$7.0~keV (iron line) band. 
The dotted lines are the Galactic coordinate with the grid spacing of 5$'$. 
The source at the lower-right is {\axj} and the spectrum is taken from 
the  solid circle of a 3$'$-radius.  Background spectrum is taken from the annulus defined by the solid and dashed (radius = 4$'$) circles.  
The brightest source in the upper-left is the Sgr B2 cloud.}
\end{figure}

\begin{figure}
\psfig{file=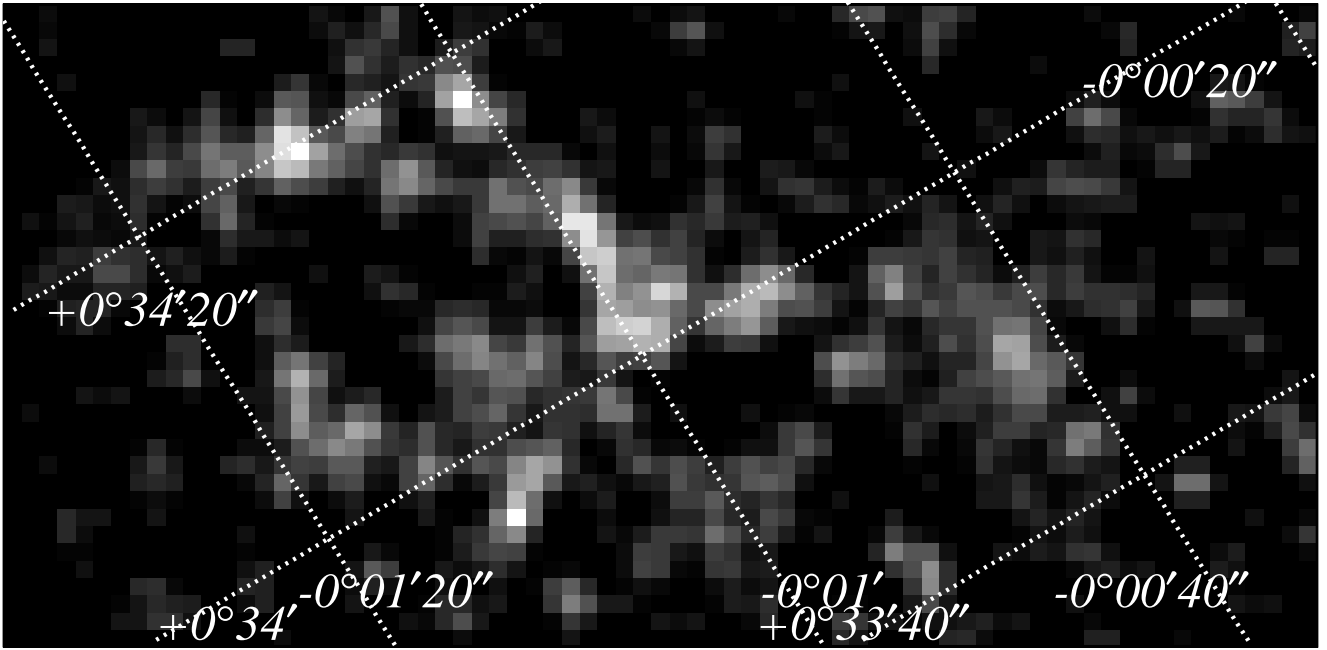,width=\textwidth}
\figcaption[f2.eps]{{\chandra}~ACIS-I image of {\axj}~in the 4.0$-$7.0 keV band.
 The dotted lines are the Galactic coordinate with the grid spacing of 20$''$.
{\axj}~is appeared to be a 10$''$ radius ring with an east-west tail.
This image is smoothed with a Gaussian of $\sigma$ = 2 pixels. 
Before the smoothing, the brightest pixel has 3 photons.
The brightness increases with a linear scale from 0.1 (black pixel) to 1.3 photons (white pixel).
}
\end{figure}

\begin{figure}
\psfig{file=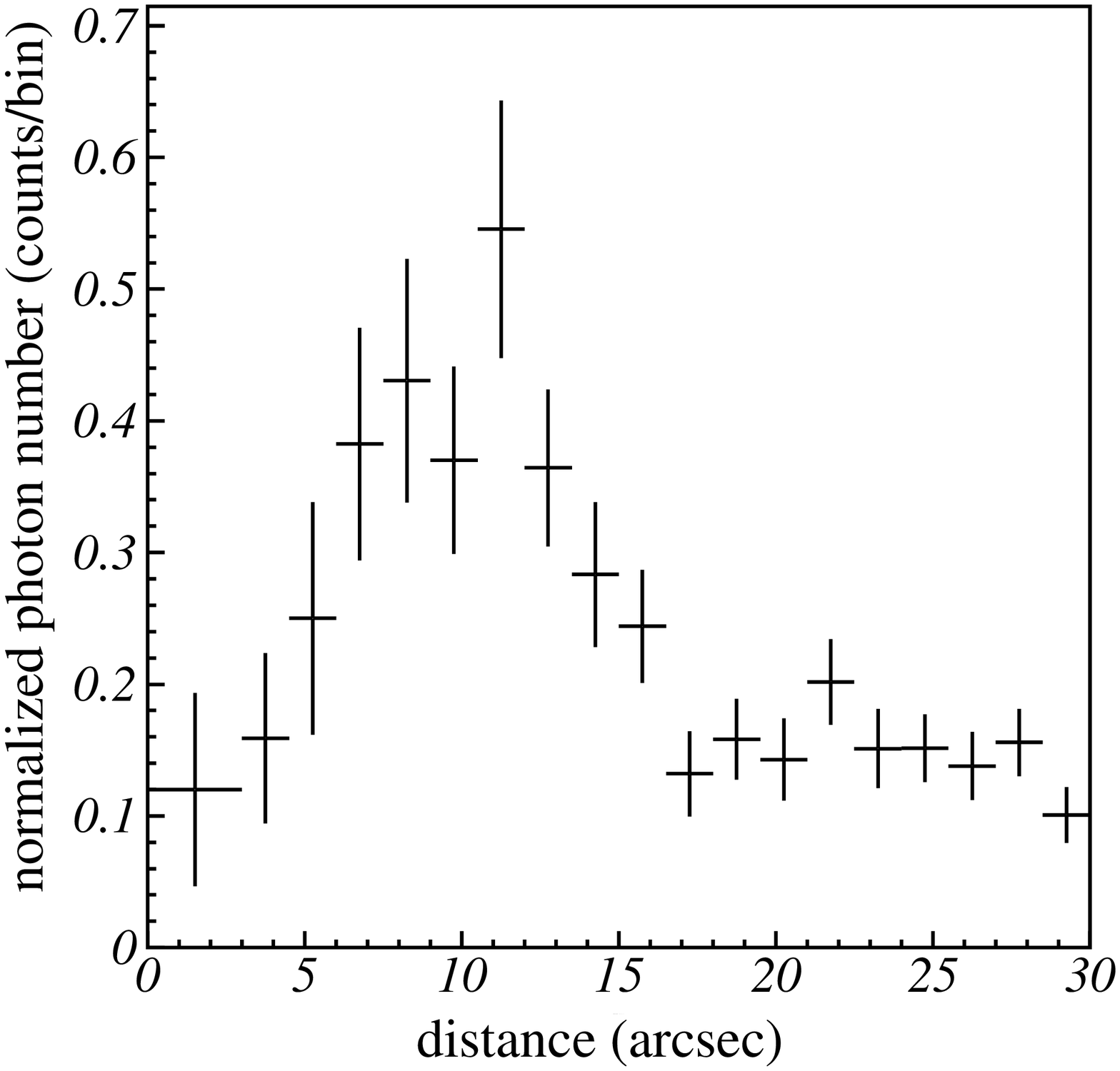,width=\textwidth}
\figcaption[f3.eps]{The radial profile of the ring structure of {\axj}.
The ring center is  (17$^{\rm h}$47$^{\rm m}$02$\fs$6,
 $-$28$\arcdeg$27$\arcmin$33$\arcsec$)$_{\rm{J}2000}$.}
\end{figure}

\begin{figure}
\psfig{file=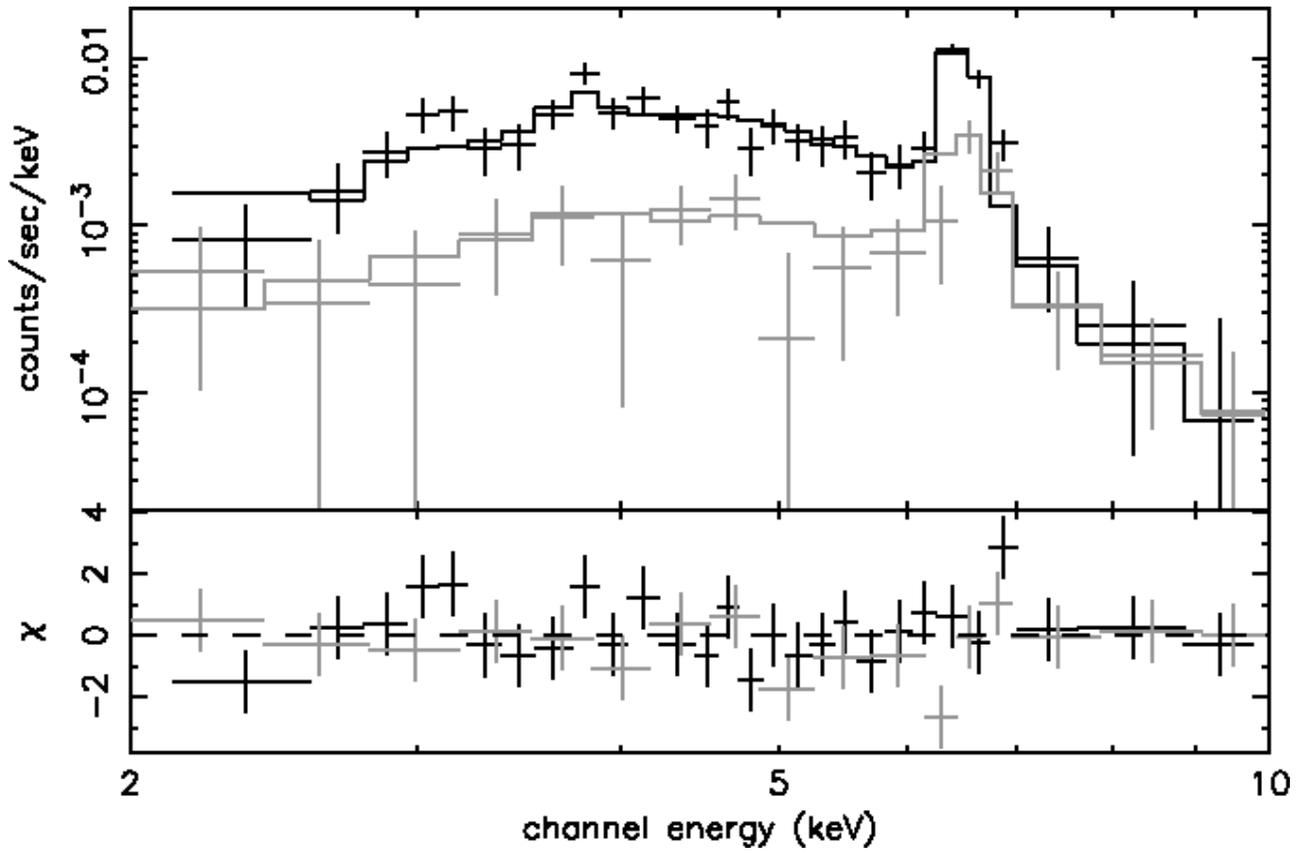,width=\textwidth}
\figcaption[f4.eps]{The X-ray spectra of {\axj}~ 
obtained from {\asca}~(gray) and {\chandra}~(black). 
The data points are given by crosses while the solid lines are 
the best-fit NEI plasma model. 
The data residuals are shown in the lower panel.}
\end{figure}

\clearpage

\scriptsize

\begin{deluxetable}{ccccccc}
\tablewidth{0pt}
\tablenum{1}
\tablecaption{ Best-fit parameters of  NEI model obtained from the combined fit of the \asca ~and \chandra ~spectra}
\tablehead{\colhead{$N_{\rm H}$\tablenotemark{a}} &\colhead{$kT$\tablenotemark{b}}
          &\colhead{$nt$\tablenotemark{c}} &\colhead{Metal abundances\tablenotemark{d}}  
          &\colhead{Flux\tablenotemark{e}}  
          &\colhead{$L$x\tablenotemark{f}}
          &\colhead{$\chi^{2}$/d.o.f} \nl
           [10$^{22}$Hcm$^{-2}$] &[keV] 
          &[10$^{10}$ s$^{-1}$ cm$^{-3}$] &[solar]   
          &[10$^{-13}$ergs cm$^{-2}$~s]
          &[10$^{34}$ergs s$^{-1}$] &
}
\startdata
13.9(10.7--17.2) & 6.1(3.1--25.8) & 1.7(1.3--2.7) & 
4.5(1.6--9.9) & 8.2 & 1.3 & 41.3/40 \nl
\enddata 

\tablecomments{Errors are at 90\% confidence level.}
\tablecomments{All parameters of the model(including normalization) are best-fitted with both {\asca} and {\chandra} data points. }
\tablenotetext{a}{Hydrogen column density.}
\tablenotetext{b}{Temperture of electrons in a thin thermal plasma.}
\tablenotetext{c}{Ionization parameter is an NEI.}
\tablenotetext{d}{Metal abundances (helium fixed at cosmic). The elements included are C, N, O, Ne, Mg, Si, S, Ca, Fe, Ni.}
\tablenotetext{e}{Observed flux in the 2.0--10.0 keV band. }
\tablenotetext{f}{Absorption-corrected luminosity in the 2.0--10.0 keV band.}

\end{deluxetable}

\begin{center}
\begin{deluxetable}{cccc}
\tablenum{2}
\tablecaption{ Comparison of the parameters between {\axj} and SN 1987A}
\tablehead{\colhead{}&\colhead{Distance}&\colhead{radius of an X-ray shell}
	  &\colhead{$t$} \nl
	  &[kpc] &[pc] &[years]}

\startdata
{\axj} &8.5 &0.4 &80  \nl
SN 1987A &50\tablenotemark{a} &0.12\tablenotemark{a} &13  \nl
\enddata

\tablenotetext{a}{The parameters of SN 1987A are taken from Burrows et al. (2000)}
\end{deluxetable}
\end{center}

\end{document}